\documentclass[manuscript]{aastex}

\shorttitle{Scatter of mass-richness relation}
\shortauthors{Campa et al.}

\begin{document}


\title{Measuring the scatter of the mass-richness relation in galaxy clusters in photometric imaging surveys by means of their correlation function }

\author{Julia Campa \altaffilmark{1,2,3}}
\email{campa@fnal.gov}
\and
\author{Brenna Flaugher\altaffilmark{1}}
\and 
\author{Juan Estrada \altaffilmark{1}}


\altaffiltext{1}{Center for Particle Astrophysics, Fermi National Accelerator Laboratory, Batavia, IL}
\altaffiltext{2}{Centro de Investigaciones Energ\'eticas, Medioambientales y Tecnol\'ogicas (CIEMAT), Madrid, Spain}
\altaffiltext{3}{Universidad Autonoma de Barcelona (UAB), Bellaterra, Barcelona, Spain}


\begin{abstract}
Knowledge of the scatter in the mass-observable relation is a key ingredient for a cosmological analysis based on galaxy clusters in a photometric survey. In this paper we aim to quantify the capability of the correlation function of galaxy cluster to constrain the intrinsic scatter $\sigma_{\ln M}$. We demonstrate how the linear bias measured in the correlation function of  clusters can be used to determine the value of this parameter. The new method is tested in simulations of a $5,000$ $deg^2$ optical survey up to $z\sim1$, similar to the ongoing Dark Energy Survey (DES). Our  results show that our method works better at lower scatter values. We can measured the intrinsic scatter $\sigma_{\ln M}=0.1$ with a standard deviation of $\sigma(\sigma_{\ln M})\sim0.03 $ using this technique. However,  the expected intrinsic scatter of the DES RedMaPPer cluster catalog $\sigma_{\ln M}\sim 0.2$ cannot be recovered with this method at suitable accuracy and precision because the area coverage is insufficient. For future photometric surveys with a larger area such as LSST and Euclid, the statistical errors will be reduced. Therefore, we forecast higher precision to measure the intrinsic scatter including the value mention before. We conclude that this method can be used as an internal consistency check method on their simplifying assumptions and complementary to cross-calibration techniques in multi-wavelength cluster observations.
\end{abstract}





\section{Introduction}
The discovery of late time cosmic acceleration from observations of supernovae in 1998 is one of the most important development of modern cosmology (\citet{1998AJ....116.1009R}; \citet{1999ApJ...517..565P}). It raises fundamental questions about the expanding universe and our understanding of gravity. 
The cosmic acceleration could arise from the repulsive gravity of dark energy or it may be signal that General Relativity breaks down on cosmological scales and must be replaced (e.g.,  \citet{2006IJMPD..15.1753C}; \citet{2012PhR...513....1C}). Growth of structures methods using galaxy cluster surveys can test these theories at recent epoch (e.g., \citet{2013PhR...530...87W}).


Clusters of galaxies were first identified as over-dense regions in the projected number counts of galaxies (e.g., \citet{1958ApJS....3..211A}; \citet{1968cgcg.book.....Z}). They are the most virialized systems known in the Universe and have a long history as cosmological probes. 

The abundance of galaxy clusters as a function of mass can be used to constrain cosmological parameters (e.g., \citet{1993MNRAS.262.1023W}; \citet{2010ApJ...708..645R}; \citet{2011ARA&A..49..409A};\citet{2015arXiv150207357B} )  and they are also a powerful tool for large scale studies (e.g., \citet{1998lsst.conf..137B}; \citet{2003ApJ...599..814B}; \citet{2001NewAR..45..355E}; \citet{2005MNRAS.357..608Y}; \citet{2008ApJ...676..206P};  \citet{2013MNRAS.430..134W}). 

The quantity most tightly constrained by cluster abundance is a combination of the form $\sigma_8\Omega_m^q$. However, the statistical power of large solid angle cluster surveys will allow us to break the degeneracy between $\sigma_8$  and $\Omega_m$.  The evolution of cluster abundance with redshift is highly sensitive to cosmology because the matter density, $\Omega_m$ controls the rate at which structure grows.  The evolution of the abundance will also allow us to constrain the equation of state, $w$ (e.g., \citet{1998MNRAS.298.1145E}; \citet{2001ApJ...553..545H}; \citet{2005ASPC..339..140M}) and to parameterize deviations from General Relativity (e.g., \citet{2015PhRvD..92d4009C}).

The number density of virialized dark matter halos as a function of  redshift and halo mass can be accurately predicted from N-body simulations (e.g., \citet{1999MNRAS.308..119S}; \citet{2001MNRAS.321..372J}; \cite{2003MNRAS.346..565R}; \citet{2006ApJ...646..881W}; \citet{2007ApJ...671.1160L}; \citet{2008ApJ...688..709T}; \citet{2010MNRAS.403.1353C}; \citet{2012MNRAS.426.2046A}; \citet{2013MNRAS.433.1230W}). Comparing these predictions to the evolution of the abundance of galaxy clusters in large-area  surveys that extend to high redshift ($z \geq 1$) can provide precise constraints on the cosmological parameters. 

Massive galaxy clusters can be identified via optical (e.g., \citet{2011ApJ...727...45S}; \citet{2014ApJ...785..104R}), X-ray emission (e.g., \citet{1998ApJ...502..558V}; {\citet{2000ApJS..129..435B};\citet{2007MNRAS.382.1289P}) and Sunyaev-Zeldovich effect (SZE) (e.g.,  \citet{2014A&A...571A..29P}; \citet{2015ApJS..216...27B}) observables. Their masses can be estimated in a number of different ways using these detections. However, these estimators are always indirect and inferred from observables correlated with mass. Since the number density of clusters is a strong function of mass, a well-understood mass-observable relation is required to recover the cosmological information. 
Uncertainties in the mass-observable relation are the main challenge for cosmological studies with clusters, and could destroy most of the cosmological information from cluster counts if it is not well calibrated (e.g., \citet{2005PhRvD..72d3006L}). The calibration task is to determine the mean relation and the dispersion of the mass-observable relation (called ``scatter'').

There is a long history of cluster samples selected from optical and near infrared photometric surveys (e.g., \citet{2000AJ....120.2148G}; \citet{2007ApJ...660..239K}; \citet{2010ApJS..191..254H}; \citet{2014ApJ...785..104R}; \citet{2015ApJS..216...20B}), and large scale optical surveys will soon be available from ongoing and future surveys such as the Dark Energy Survey (DES, \citet{2005IJMPA..20.3121F})Ê\footnote{http://www.darkenergysurvey.org}, Euclid \citep{2011arXiv1110.3193L} and the Large Synoptic Survey Telescope (LSST; \citet{2012arXiv1211.0310L}). They are expected to generate galaxy catalogs to sufficient depth to reliably detect clusters at redshift} as high as $z\sim1$.


In order to overcome the degeneracy between cosmological parameters and mass calibration parameters, self-calibration techniques have been developed (e.g., \citet{2003A&A...398..867S}, \citet{2004ApJ...613...41M}, \citet{2005PhRvD..72d3006L}, \citet{2011PhRvD..83b3008O} and \citet{2012A&A...547A.117A}). The relation is calibrated using a large cluster sample complemented with statistical properties of the cluster that are sensitive to mass.  One parameterizes the mass-observable relation and then use standard likelihood methods to jointly fit for both cosmology and mass-observable parameters.  

Uncertainty in the scatter for the mass-observable relation translates into a systematic uncertainty in the determination of cosmological parameters. This systematic effect has been studied for constraining dark energy parameters in a cluster counting experiment for an imaging survey with an area of $5000$ $deg^2$, similar to the ongoing Dark Energy Survey. 
{\citet{2011ApJ...740...53R}  studied when a source of scatter is observationally relevant with the standard calibration technique.
 Their conclusions are that if the accuracy to measure the scatter, $\sigma_{\ln M} - \sigma_{\ln M}^{true}$, is  $\geq 0.05$, the recovered dark energy parameters will be significantly biased and the source of noise will be observationally relevant. 

In this paper, we present a new method to constrain the scatter of the mass-observable relation for ongoing and future wide area photometric surveys. We show that the amplitude of the correlation function of clusters 
(i.e., \citet{2003ApJ...599..814B}, \citet{1983ApJ...270...20B}, \citet{2009ApJ...692..265E})  provides information about the mass-observable relation \citep{2004ApJ...613...41M}, and can be used to constrain the scatter. 
This method is complementary to self-calibration and cross-calibration techniques in multi-wavelength data from so called direct mass measurement (e.g., \citet{2009ApJ...699..768R} and \citet{2015MNRAS.454.2305S}). 

A basic step towards the analysis of the clustering of clusters is the understanding of the dark matter halos. First, we make use of N-body simulations with $5,000$ $deg^2$ to study the accuracy of the halo abundance and the halo-mass bias. We compare our measurements with the available fitting formulae. Then we construct the cluster catalogs using this halo mock light cone simulation. We assign richness to the dark matter halos by means of a mass-richness relation and study how accurately we will measure the scatter in photometric surveys with a DES volume. In this work, we assess if the new analysis technique to constrain the scatter can be implemented as a cross-check of the methods mentioned before in the DES data and other  photometric surveys.
 

The plan of this paper is the following. In Section 2 we describe the mass-richness relation that we are going to use in this work to add the galaxies to the dark matter halo in simulations. In Section 3, we will use the halo model to predict cluster clustering and construct a theoretical model for the bias to compare with observations in optical surveys such as DES. 
  In Section 4, previous to studying the bias in clusters, we study the accuracy of the halo bias and mass function models by comparing them with results from dark matter halo numerical simulations. In Section 5, we describe our simulated cluster sample based on the dark matter halo simulations. In Section 6, we define our model for the bias when we make cuts in richness to compare with the measurements in simulations. In Section 7  we will show how measurements of the clustering of clusters can constrain the scatter of the scaling relation. We will make a forecast of the performance of the new analysis technique for DES. We end with a summary and conclusions in Section 8. 


\section{Mass-richness relation of galaxy clusters}
\label{sec:m-N}
The main observational challenges when using clusters to constrain cosmology are the cluster detection algorithm and the cluster mass estimation. The advent of multi-band data have led to a proliferation of optical cluster-finding algorithms. 
These cluster finders  estimate a richness that correlates with external mass proxies and then the mass-richness relation can be calibrated. The algorithms use various techniques for measuring clustering in angular position plus color/redshift space, ranging from simple matched filters to Voronoi tessellations. Some examples of the matched filter algorithms that use the red-sequence are the MaxBCG \citep{2007ApJ...660..221K} and the more modern redMaPPer  \citep{2014ApJ...785..104R}.  The Voronoi Tesselation (VT; \citet{2011ApJ...727...45S}) algorithm uses photometric redshift to detect clusters in 2+1 dimensions.  

Although the mass-richness relation  is being calibrated using many cluster finder algorithms such as redMaPPer  and the VT algorithms in the DES footprint (e.g., \citet{2015MNRAS.454.2305S}), we are going to use the form of the mean relation between the cluster mass and richness used in \citet{2009ApJ...699..768R} to test our method in simulations. It is based on the results from the statistical weak lensing analysis in the MaxBCG cluster catalog  \citep{2007ApJ...660..239K}. This algorithm identifies  clusters using two optical properties. First, the brightest cluster galaxy (BCG) typically lies near the center of the cluster galaxy light distribution. Second, the cores of rich clusters are dominated by red-sequence galaxies that occupy a narrow locus in color-magnitude space, the E/SO ridge line. MaxBCG uses a maximum-likelihood method to evaluate the probability that a given galaxy is a BCG near the center of a red-sequence galaxy density excess.  

Every cluster is also assigned a richness measure $N_{200}$, which is the number of red sequence galaxies above a luminosity cut of $0.4L_*$ and within a specified scaled aperture, centered on the Brightest Cluster Galaxy (BCG) of each cluster. 

The separate statistical weak lensing measurements of \citet{2007arXiv0709.1159J} and \citet{2008MNRAS.386..781M} indicate that $N_{200}$ is strongly correlated with cluster virial mass. These analyses are discussed in the Appendix of \citet{2009ApJ...699..768R} and yield a relation between cluster mass and richness given by,
\begin{equation}
\frac{\langle M | N_{200} \rangle}{10^{14}}=\exp^{B_{M|N_{200}}}\Bigr(\frac{N_{200}
}{40}\Bigl)^{\alpha_{M|N_{200}}},
\label{eq:M_N_relation}
\end{equation}
where $\alpha_{M|N}=1.06\pm 0.08 (stat) \pm 0.23 (sys)$ and $B_{M|N}=0.95  \pm 0.07 (stat) \pm 0.10 (sys)$  are the priors described in \citet{2009ApJ...699..768R}. We take them as the fiducial values of the mass-richness relation parameters.

\section{A model for the galaxy cluster correlation function}
\label{sec:galaxy_cluster_correlation_function}
In this section the clustering of clusters 
is predicted using the halo model of galaxy clustering (i.e. \citet{2002PhR...372....1C}). We require a Halo Occupation Distribution (HOD) where the mean number of galaxies is specified, and the spatial and velocity distributions of galaxies within halos (e.g., \citet{1952ApJ...116..144N}; \citet{2002ApJ...575..587B}; \citet{2013PASA...30...30B}; \citet{2014MNRAS.442.1930P}).
One can then calculate the clustering of clusters from the combination of the HOD with the clustering of halos if we assume that the clustering of halos depends only of the halo mass. There are discussions about the dependency of the HOD on the cosmic environment in addition to the mass of the halo (e.g., \citet{2012MNRAS.425.2766C}), however we develop the formalism of the halo model keeping it as simple as possible.
 
From the assumption that all galaxies reside within dark matter halos it follows immediately that given a halo population and a HOD, we can calculate the correlation function of clusters. This is written as the sum of the one halo term and the two halo term.  On large scales the two halo term dominates the correlation function and it can be expressed in terms of the weighted value of the halo bias. Thus, the galaxy cluster correlation function simplifies to 
 \begin{equation}
\xi_{cc}(\bar{r},z)=b^2(z)\xi_{mm}(\bar{r},z),
\end{equation}
where  $b(z)$ is the mean large scale bias of a particular galaxy population at redshift $z$ that we assume is constant at large scales. 
The dark matter correlation function, $\xi_{mm}$, is obtained via Fourier transform of the non-linear dark matter power spectrum, $P_{NL}$. In three dimensions, after assuming space isotropy,  this yields 
\begin{equation}
\xi_{mm}(r)=\frac{4\pi}{(2\pi)^3} \int P_{NL}\frac{sin(kr)}{kr} k^2 dk .
\end{equation}
For the $\Lambda CDM$ model parameters, the predicted non-linear dark matter model is calculated using the non-linear Halo-fit power spectrum (\citet{2003MNRAS.341.1311S}; \citet{2012ApJ...761..152T}). 

In terms of the halo mass function, $\frac{dn(M,z)}{d \ln M}$, and the linear halo bias, $b(M,z)$, 
the mean large scale bias is given by 

\begin{equation}
b(z)=\frac{1}{\bar{n}}\int d \ln M \frac{dn(M,z)}{d \ln M}\langle N| M \rangle b(M,z),
\label{eq:b_modelo}
\end{equation}
where $\langle N| M \rangle$ is the mean number of galaxies per halo and $\bar{n}$ is the mean number density of galaxies given by
\begin{equation}
\bar{n}=\int dM \frac{dn(M,z)}{d \ln M}\langle N| M \rangle.
\end{equation}

\section{Halo mass function and bias}
\label{sec:mf}

In this section we study how the dark matter halos are biased with respect to the underlying matter distribution using the halo model. We study the accuracy of the \citet{1999MNRAS.308..119S} and  \citet{2010ApJ...724..878T} halo bias  theoretical models using dark matter simulations. We use a halo catalog with the same volume as DES based on the Hubble Volume PO light cone output (\citet{2002ApJ...573....7E}), extracted from the DES\_v1.02 mock galaxy catalog\footnote{http://www.slac.stanford.edu/mbusha/mocks/catalogs.html; provided by M. Busha \& R. Wechsler}. The halos were identified directly on the dark matter light cone using a spherical over density halo finder.  The algorithm defines spherical regions that are overdense with respect to the critical density $\rho_c$.} 

The halos mass $M_{200}$ is defined as the mass enclosed in a sphere of radius $R_{\Delta}$ whose mean density is $\Delta=200$ times the threshold density. So the $M_{200}$ mass is given by,
 \begin{equation}
 M_{200}=\frac{4}{3}\pi \Delta \rho_c R_{200}^3.
 \label{eq:M_200}
 \end{equation}

\subsection{Mass function definition and results in simulations}

Since the theoretical models for the halo bias have been derived from the mass function (e.g.  \citet{1996MNRAS.282..347M} and \citet{1999MNRAS.308..119S}), we first study the theoretical models of the halo abundance.
The comoving number density of halos with mass between $M$ and $M+dM$ or the unconditional mass function can be written as 
\begin{equation}
\frac{dn}{dM}= \frac{\bar{\rho}_m}{M}f(\nu)\frac{d\nu}{dM},
\label{eq:mf}
\end{equation}
where $f(\nu)$ is the multiplicity function (the fraction of mass in collapsed objects) and  $\bar{\rho}_m$ is the mean comoving mass density. 
The height of the density peaks is defined 
\begin{equation}
\nu \equiv \frac{\delta_c^2}{\sigma^2(M)},
\end{equation}
where $\delta_c=1.686$ is the critical density for spherical collapse and $\sigma^2(M)$ is the variance of matter density fluctuations on mass scale $M$. 

\citet{1999MNRAS.308..119S} generalized the expression of the Press-Schechter mass function (\citet{1974ApJ...187..425P}) and calibrated the free parameters using numerical simulations. It can be written as
\begin{equation}
\nu f(\nu)=A(p)\sqrt{\frac{q\nu}{2\pi}}[1+(q\nu)^{-p}] e ^{-\frac{q\nu}{2}},
\end{equation}
with $p=0.3$ and $q=0.707$ and $A(p=0.3)=0.322$.

Later, \citet{2008ApJ...688..709T} also calibrate fitting functions for the mass function and bias using high resolution simulations. They choose the form
\begin{equation}
\nu f(\nu)=A[1+(b\nu)^a]\nu^d e^{-\frac{c\nu}{2}},
\end{equation}
where $A$, $a$, $b$, $c$ and $d$ are the free parameters for each overdensity $\Delta$ value with respect to the mean density of the universe, $\bar{\rho}_m$. These parameters were calibrated in simulations at $z=0$. 
They also provide redshift correction to match mass function to simulations.

In the DES light cone, the halo finder defines overdense regions with respect to the critical density $\rho_c(z)$ instead of $\bar{\rho}_m$. So If we define an overdensity contrast as , $\Delta'=\frac{\Delta}{\Omega_m(z)}$, we can use this functional form for any value of $\Delta'$.  The value of the parameters at $z=0$  are calculated by spline interpolation as a function of $\Delta'$ and then we calculate their redshift evolution.

\begin{figure}
\includegraphics[scale=0.9]{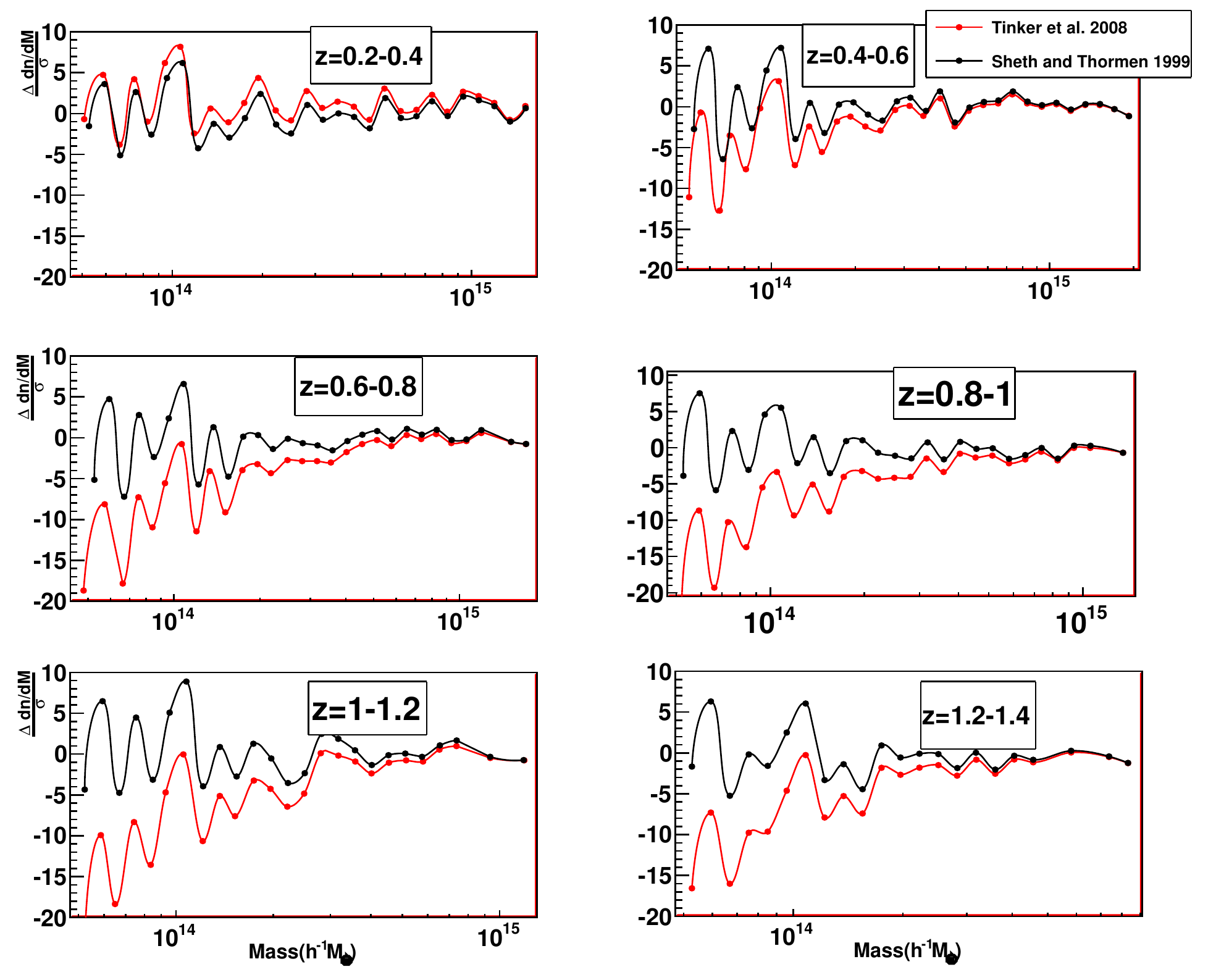}
\caption{Systematic error $\Delta \frac{dn}{dM}$ compared with the statistical error $\sigma$ for the six redshift bins from the light cone simulations. Black dots are the values with the \citet{1999MNRAS.308..119S} model with the best p and q values and red dots are the \citet{2008ApJ...688..709T}  model.  \label{fig:fig0}}
\end{figure}

We compare the mass function measured in redshift bins of width $\Delta z=0.2$ using the dark matter halo simulation with a DES volume with  the \citet{1999MNRAS.308..119S} model with $p$ and $q$ fiducial values evaluated at the mean redshift. Since there is a high disagreement in all the mass ranges, we fit the parametric model to the halo catalog measurement. Our fitting method is a simple $\chi^2$ of the difference between the theoretical model and the measured counts in bins (e.g., \citet{2001MNRAS.321..372J}, \citet{2010MNRAS.402..589M} and \citet{2011MNRAS.415..383M}). We also studied the accuracy of the fitting function for $\Delta$ overdensities of  \citet{2008ApJ...688..709T}.  Figure \ref{fig:fig0} shows the comparison of the systematic error $\Delta \frac{dn}{dM}$  with the statistical error $\sigma$ for the two models. In all the redshift bins we found that the deviations, $ \frac{\Delta \frac{dn}{dM}}{\frac{dn}{dM}}$,  increase on the high mass tail for both models where the number of halos is very small. However, these deviations are not significant.  

The results also show that the disagreement between the two models increases with redshift. We found better accuracy in all the redshift bins with the  \citet{1999MNRAS.308..119S} parameters fitted by us.  But, of course, one can also fit  the Tinker parameters to simulations instead of doing an interpolation and compare again these models.  We postpone this work to the future.
\subsection{Halo bias definition and results in simulations}
\begin{figure}
\includegraphics[scale=0.6]{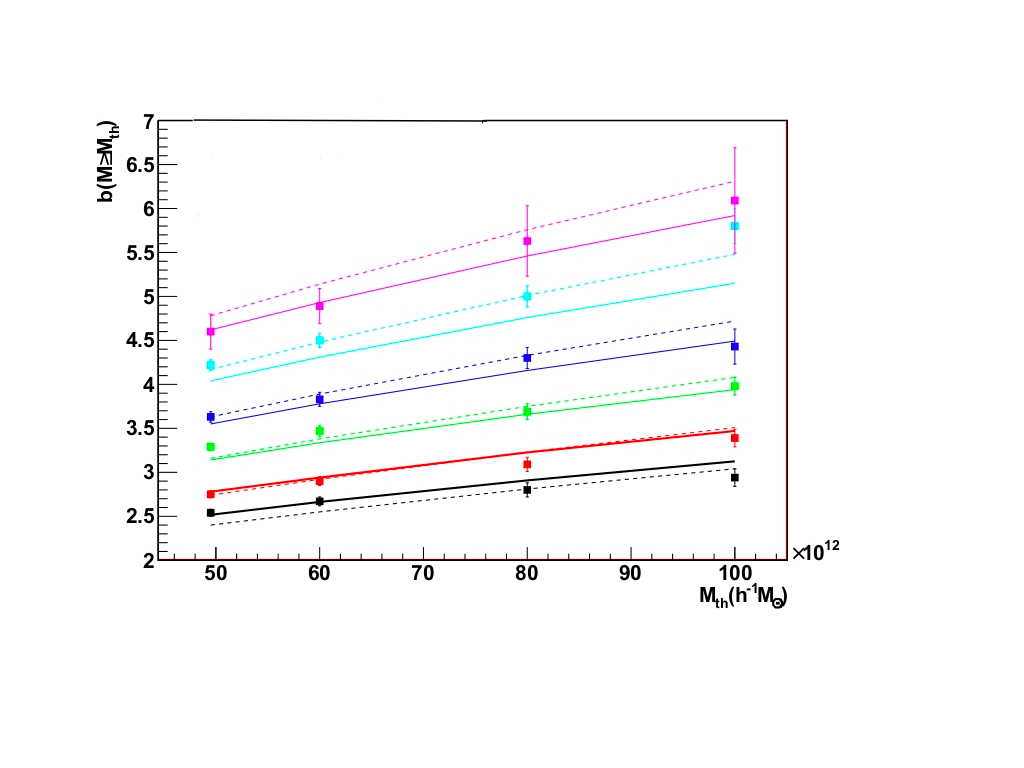}
\caption{Comparison of the halo bias model with the measurements in the light cone for the six redshift bins $z=0.2-0.4$, $z=0.4-0.6$, $z=0.6-0.8$, $z=0.8-1$, $z=1-1.2$ (black, red, green, blue,cyan and pinks dots respectively).  The solid curves are the values of  the  \citet{1999MNRAS.308..119S} model with the best p and q values and the dashed curves the  \citet{2010ApJ...724..878T}  model. The random catalog is 5 times denser than the catalog, $N_R=5N_D$ to optimize the Poisson noise. \label{fig:fig00}}
\end{figure}

The corresponding large scale halo bias prediction of \citet{1999MNRAS.308..119S} is given by
\begin{equation}
b(\nu)=1+\frac{q\nu-1}{\delta_c} + \frac{2p}{\delta_c(1+(q\nu)^p)}.
\end{equation}
Later, \citet{2010ApJ...724..878T} introduces a similar but more flexible fitting function of the form 
\begin{equation}
b(\nu)=1-A\frac{\nu^a}{\nu^a+\delta_c^a}+B\nu^b+C\nu^C,
\end{equation}
where the parameters also depend on the density contrast $\Delta$. 
Using the measurements of the halo correlation function for four mass thresholds and six redshift bins of with $\Delta z=0.2$ of the light cone simulations, we measure the halo linear bias where it is considered to be deterministic and scale independent. For each redshift bin, we fit the matter correlation function $\xi_{mm}(r)$ at a given $z$ to the one measured in  the simulations $\xi_{hh}$ using the \citet{1993ApJ...412...64L} estimator.  We optimized the Poisson error and made an estimation of the cosmic variance using the jackknife method. Then we compare these measurements with the predictions of the \citet{1999MNRAS.308..119S} and the \citet{2010ApJ...724..878T} bias models as shown in Figure \ref{fig:fig00}. We consider the difference between these two models as a systematic uncertainty of our method. We note also that the bias errors increase with increasing mass and redshift because the number of halos is lower.

\section{Construction of the cluster catalog with a DES volume}
\label{sec:simulation}
We created cluster catalogs using the DESv1.02 halo mock catalog light cone mention before. 
The dark matter halos of this simulation are populated with galaxies using a model of HOD. We assign a richness $N$ to dark matter halos by means of a conditional distribution $P(N|\ln M)$ for a halo of  mass $M$ to contain $N$ galaxies. 
 


As discussed in 
\citet{2002A&A...383..773I}, \citet{2005PhRvD..72d3006L} and \citet{2011PhRvD..83b3008O}, we assumed the intrinsic scatter $\sigma_{\ln M}$ in the scaling relation to be log-normally distributed around the mean scaling relations, i.e., Gaussian or normal in $\ln M$. Thus, the probability of observing the richness $N$ given the true underlying mass $M$ is given by
\begin{equation}
P(N| \ln M)=\frac{1}{\sqrt{2\pi \sigma_{\ln M}^2}} \exp \Bigl[-\frac{1}{2\sigma_{\ln M}^2}(\ln \langle M|N\rangle-\ln M)^2\Bigr].
\label{eq:P_N}
\end{equation}

In our case, the underlying mass $M$ is the halo mass $M_{200}$ in the light cone simulations given by Equation \ref{eq:M_200} and the mass-richness relation is given by Equation \ref{eq:M_N_relation}.

First we assume that the scatter does not  vary neither with redshift or mass and create three cluster catalogs with three intrinsic scatter $\sigma_{\ln M}=0.1, 0.2$ and $0.4$. Since the assumption of a constant scatter is almost certainly incorrect, then we include a Poisson component associated with the number of galaxies in a halo at fixed mass such as used in \citet{2015MNRAS.454.2305S}.  We convolve the probability distribution $p(N|\ln M)$ with a second log-normal distribution of variance $\sigma^2=\exp(-{\langle N | M \rangle})$, such as these two scatters add in quadrature.  Therefore, the effective scatter is given by
\begin{equation}
\sigma^2=\exp(-\ln \langle N | M \rangle)+\sigma_{\ln M}^2.
\label{eq:addcuadrature}
\end{equation}

\begin{figure}
\center
\includegraphics[scale=0.5,angle=0]{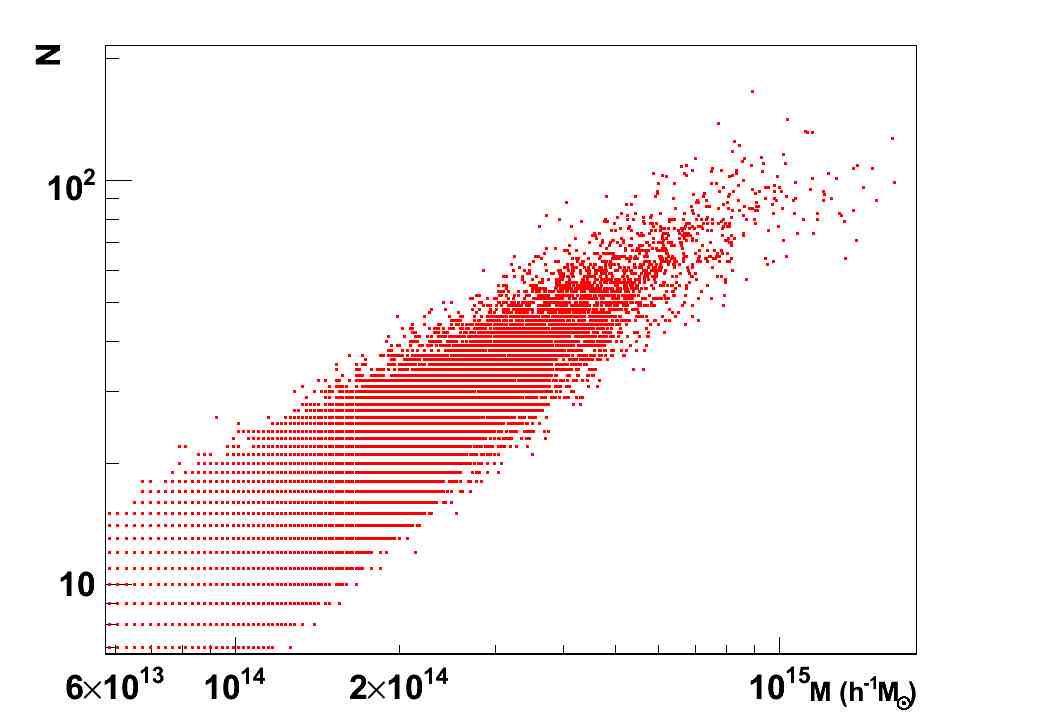}
\caption{
Red points shows the richness-mass relation when the halos of the light cone simulation are populated using a HOD distribution with an effective scatter including an intrinsic scatter, $\sigma_{lnM}^{true}=0.2$ and a Poisson noise term.}
\label{fig:fig2}
\end{figure}

\begin{figure}
\center
\includegraphics[scale=0.75,angle=90]{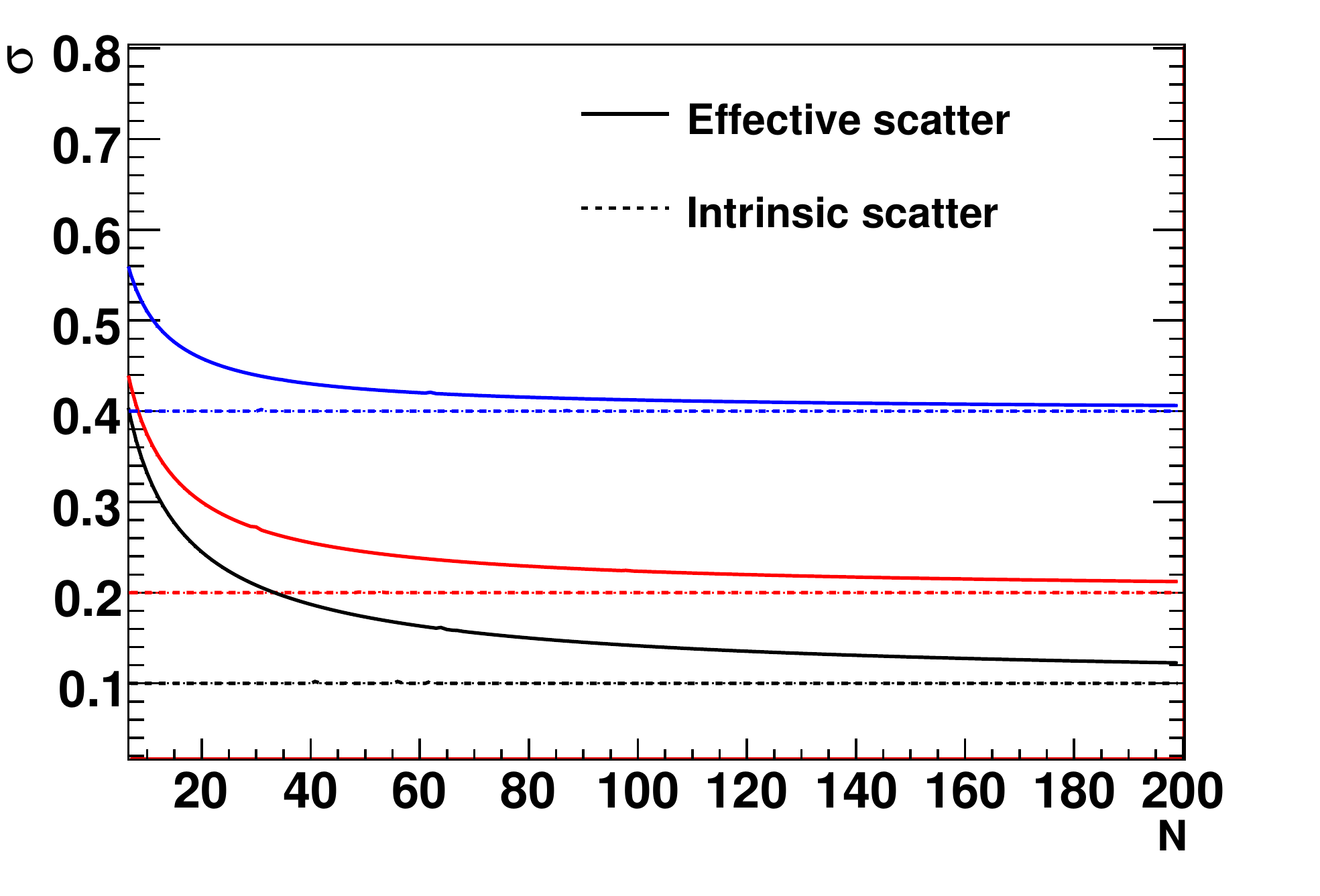}
\caption{Variation of the effective scatter  $\sigma$ with the richness (solid lines) for three intrinsic scatter values $\sigma_{\ln M}=0.1, 0.2$ and $0.4$ (black, red and blue respectivley). The comparison of the effective scatter $\sigma$  with the intrinsic scatter $\sigma_{\ln M}$ (dashed lines) is also shown. The Poisson noise term dominates at the low-mass range.  
}
\label{fig:scatter_N}
\end{figure}

Figure \ref{fig:fig2} shows an example of the mass-richness relation with a HOD distribution, $P(N|\ln M)$, with an effective scatter $\sigma$ including an intrinsic scatter $\sigma_{\ln M}=0.2$ and the Poisson noise term. Each point represents the number of galaxies that occupy a particular dark matter halo showing that the observable is noisy. 
Figure \ref{fig:scatter_N} shows the effective scatter as a function of richness for three intrinsic scatter values. The Poisson term dominates  at  the low-mass range, $M \sim 10^{14} M_{\odot}/h$ and this effect is more significant at lower scatter values. This is very important because our method may be less sensitive to uncertainties in the low-mass end and it may also work better at lower scatter values. 

Although we assume that the intrinsic scatter is constant in all the redshift slices, there is also the possibility that the scatter evolves with redshift.  In previous studies, \citet{2010PhRvD..81h3509C}  and \citet{2011PhRvD..83b3008O} model a cubic polynomial evolution.  We postpone a carefully study of the effect of the redshift evolution for a near future.

\section{Theoretical predictions for the richness bias using the Halo Model}
\label{sec:Bias_prediction}
In this section we define a model for the bias for a richness cut $N>N_{th}$ to compare with the measurements in simulations. This is given by,
\begin{equation}
b(N_{th},z)=\frac{\sum_{N=N_{th}}^{\infty} b(N,z)n_{meas}(N,z)}{\sum_{N=N_{th}}^{\infty} n_{meas}(N,z)},
\label{eq:b_mean}
\end{equation}
where $n_{meas}(N,z)$ is the number of halos per redshift and richness value measured in the simulations, and $b(N,z)$ is calculated using the halo model of galaxy clustering explained in Section  \ref{sec:galaxy_cluster_correlation_function}. The bias expected for a richness value is given by,

\begin{equation}
b(N,z)=\frac{1}{\bar{n}}\int d \ln M \frac{dn(M,z)}{d \ln M}P(\ln M|N) b(M,z),
\label{eq:b_N}
\end{equation}
and $\bar{n}$ is the mean number density of galaxies given by
\begin{equation}
\bar{n}=\int dM \frac{dn(M,z)}{d \ln M}P(\ln M|N),
\end{equation}
where the posterior $P(\ln M|N)$ distribution is related to the distribution used to create the richness catalog $P(N|\ln M)$, given by Equation \ref{eq:P_N}, using  Bayes' Theorem.

We are interested in studying the impact of the scatter in the bias. Since the bias is related to the mass function, the scatter effect in the halo abundance is also studied. As it is discussed in \citet{2005PhRvD..72d3006L}, the scatter in the mass-richness relation changes the shape and the amplitude of the mass function above an observable threshold  
significantly to provide an excess of clusters scattering up (at $N\leq N_{th}$) versus down (at $N\geq N_{th}$) across the threshold. The steepness of the mass function around the observable threshold determines this excess due to upscatters.  As the observable threshold reaches the exponential tail of the mass function, the excess of upscatter can become a significant fraction of the total and the richness bias decrease.  When  the Poisson term is not included and the effective scatter is constant, the larger the scatter the more the bias is decreased (see Figure \ref{fig:bias_model_cut_nocut}). Moreover, the impact of the scatter will be significantly greater at higher mass and redshift because the steepness of the mass function is larger there. In the next section, we will also study the effect  of the scatter on the bias when we add the Poisson noise (Figure \ref{fig:bias_model_cut_nocut_noise}). 

\section{Constraining the scatter of the mass-richness relation. Likelihood analysis}
We divide the catalog in redshift bins $\Delta z$ and make cuts in richness to measure the bias with the two point correlation function. Therefore, we have a set of $n$ bias measurements, $b_i^{meas}(N \geq N_{th},z)$ and their bias errors, $\sigma_{b_i}^{meas}$. We assume a model for the bias, $b^{model}(N_{th},z)$, with parameters $\theta=(\Lambda,\alpha_{M|N},B_{M|N},\sigma_{\ln M})$ using  Equations \ref{eq:b_mean} and \ref{eq:b_N}. Since our goal is to constrain the scatter, we consider a one dimensional likelihood given by the conditional probability distribution of the data, $\mathcal{L}=p(b^{meas}|\theta=\sigma_{\ln M})$
\begin{equation}
p(b^{meas}(N_{th},z);\theta)=\frac{1}{\sqrt{2\pi}\sigma_b^{meas}}exp\frac{-(b^{meas}(N_{th},z)-b^{model}(N_{th},z))^2}{2\sigma_{b^{meas}}^2},
\end{equation}
where we assume that the measurements are not correlated. Although this is not absolute correct, our future plan with data is to divide the catalog in richness bins where we have enough clusters. In this case, the measurements won`t be correlated.
For $n$ bias measurements, the likelihood is the product of the probabilities of the individual measurements
\begin{equation}
\mathcal{L(\theta)}=\prod_{i=1}^n p(b_i^{meas}(N_{th},z);\theta).
\end{equation}
Then we normalize the result to the unity.

First we assume known the $\Lambda CDM$ cosmological parameters of the simulations and we fix the mean mass-richness relation parameters, $\alpha_{M|N}=1.06$ and $B_{M|N}=0.95$.  Since these last parameters are never known perfectly, then we marginalized over them using the forecast  errors given in Section 2  and described in \citet{2009ApJ...699..768R}. Given a set of bias measurements and the Gaussian priors, $\pi(\alpha_{M|N})$ and $\pi(B_{M|N})$, we compute 
\begin{equation}
\mathcal{L(\theta)}=\int \mathcal{L}(\theta; \alpha_{M|N}, B_{M|N})\pi(\alpha_{M|N})\pi(B_{M|N}) d  \alpha d B_{M|N},
\label{eq:marginalize}
\end{equation}
and pick the maximum value of the likelihood.
\subsection{Forecast and error estimation for the scatter on DES volume}

 Before we study our method in simulations, we make a forecast of the precision that our method can achieve without including the systematics errors coming from the uncertainty in the theoretical bias and mass function models. Instead of the values measured in simulations, ${b_i}^{meas}$, we use the theory predictions  for a fiducial model with scatter $\sigma_{lnM}^{true}$. 
  We model the bias using using Equations  \ref{eq:b_mean} and  \ref{eq:b_N}  for 3 samples of richness threshold $N_{th}\geq$ $7$,$8$, $9$ at six redshifts $z=0.3, 0.5, 0.7, 0.9, 1.1, 1.3$. We assign each point an expected experimental error, $\sigma_{b_i^{meas}}$, obtained from the fits to the correlation function using the simulations.  First, we make a forecast of the precision with a constant intrinsic scatter.  Later, we add the Poisson noise term and compare the results.
 
\begin{figure}
\centering
\begin{minipage}{1\linewidth}
\includegraphics[scale=0.6,angle=90]{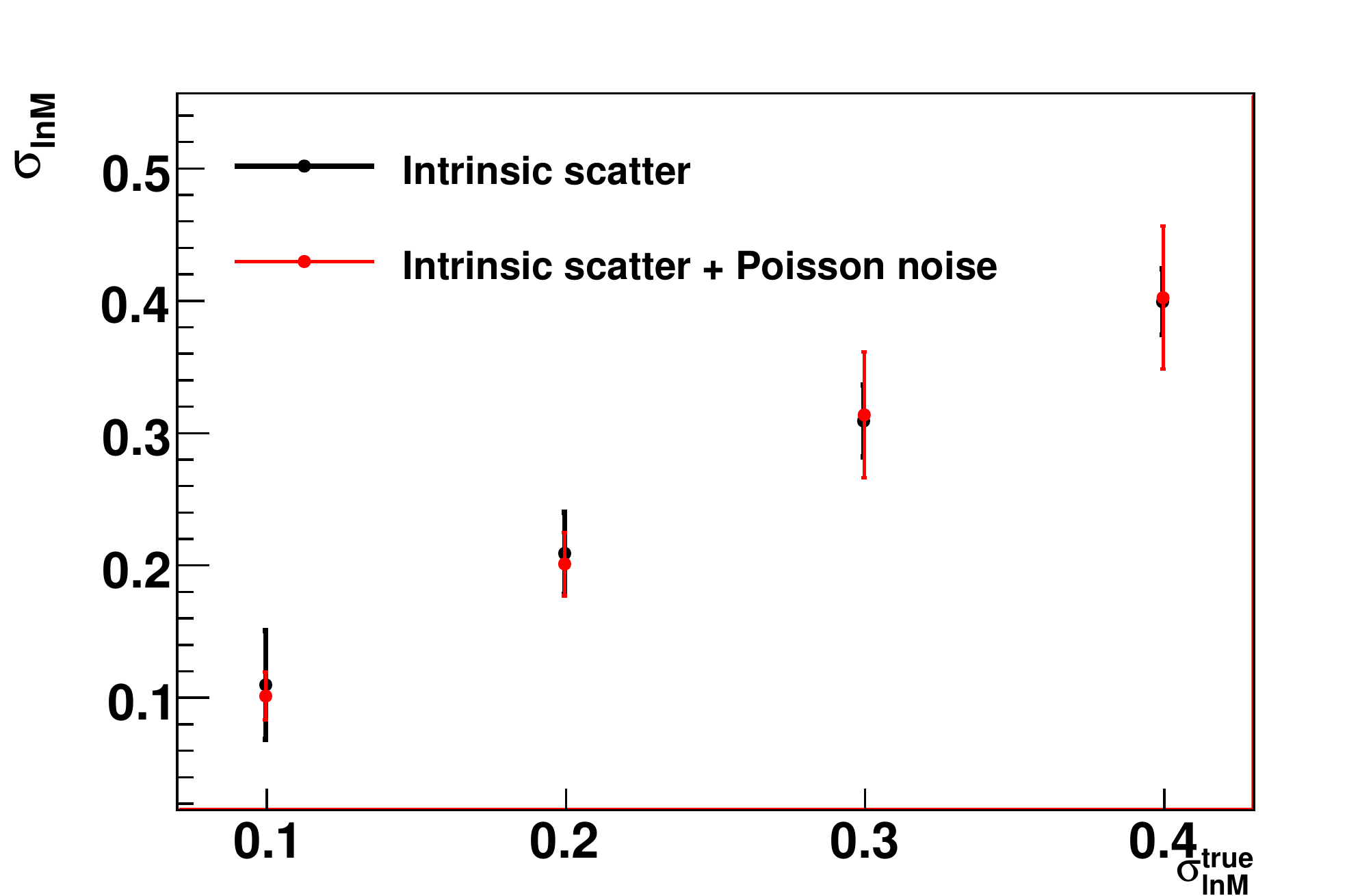}
\caption{Recovered values of the scatter, $\sigma_{\ln M}$ with their expected errors ($68 \% C.L.$) for different true scatter values, $\sigma_{\ln M}^{true}$. The comparison of the results when the effective scatter only includes the intrinsic scatter  (black dots) and when the Poisson noise (red dots) is included.}
\label{fig:fig3}
\end{minipage}
\begin{minipage}{1\linewidth}
\includegraphics[scale=0.55,angle=90]{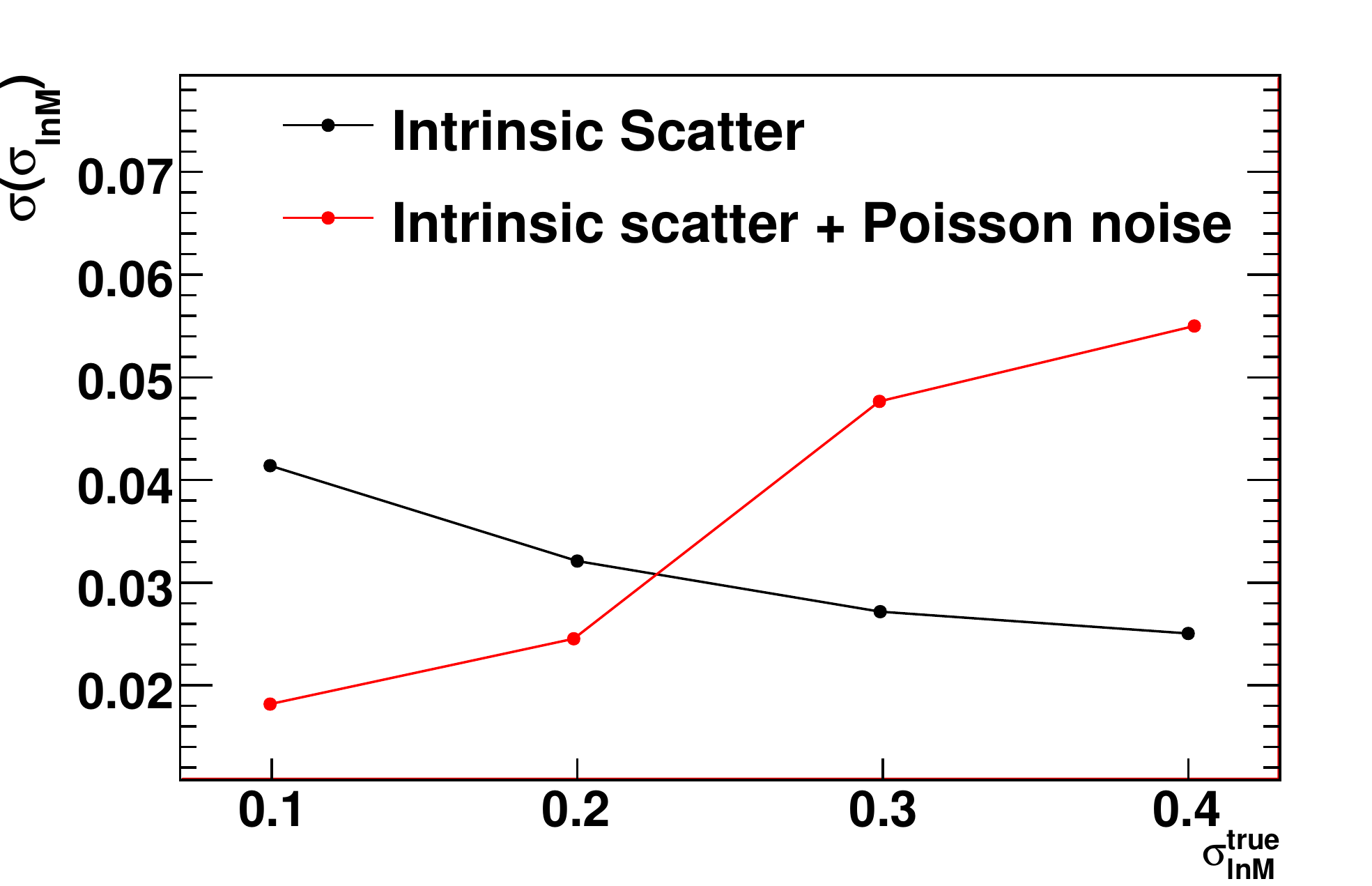}
\caption{Expected errors ($68\% C.L.$) for different scatter values taken as fiducial model, $\sigma_{lnM}^{true}$. The black dots and lines are the results when the effective scatter only includes the intrinsic scatter. The red dots and lines when the Poisson noise it is included. Our precision is better at lower intrinsic scatter values when we add the Poisson noise.}

\label{fig:fig4}
\end{minipage}

\end{figure}

\begin{figure}[htpb]
\centering
\begin{minipage}{0.49\linewidth}
\includegraphics[angle=90,width=1.\textwidth]{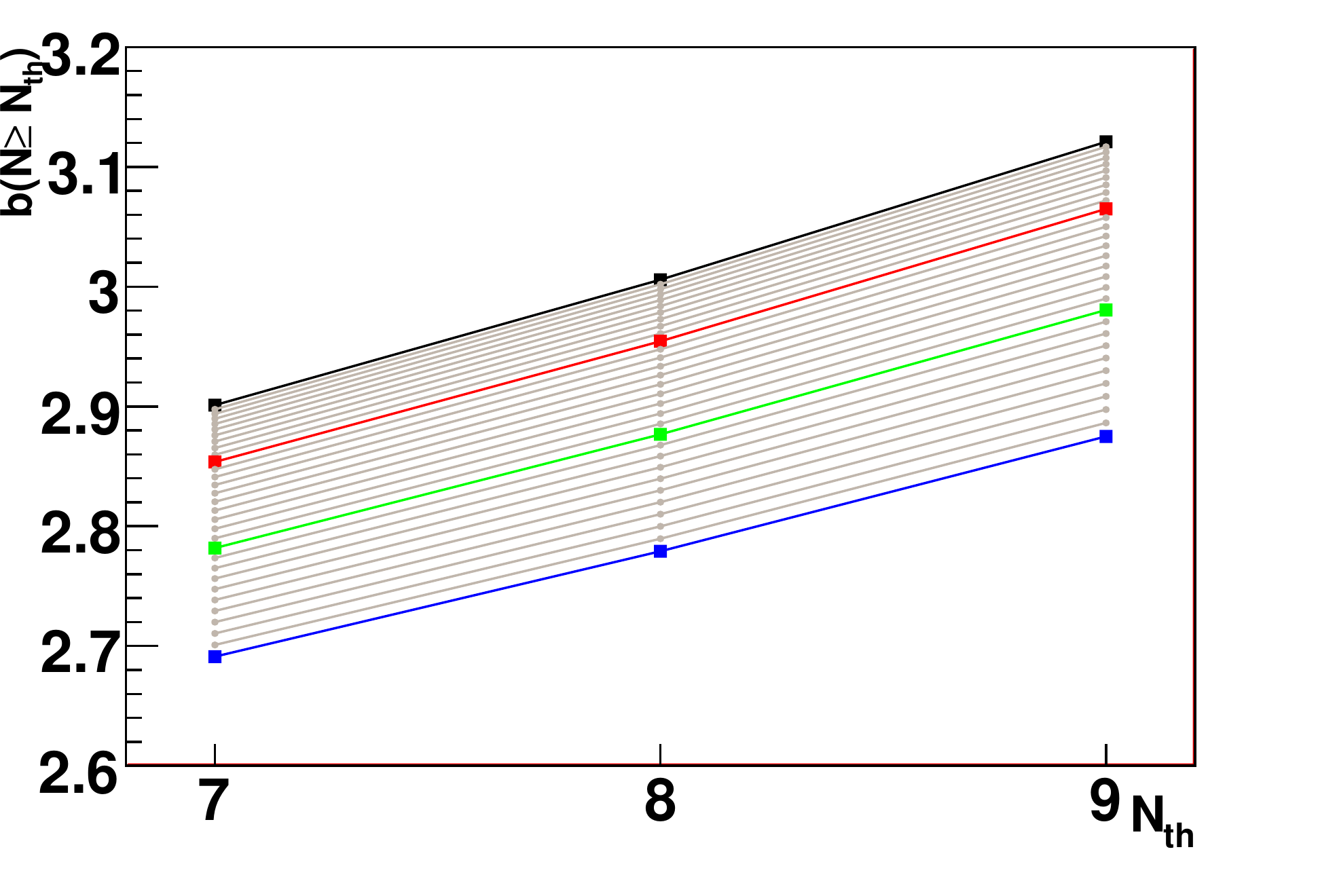}
\end{minipage}
\begin{minipage}{0.49\linewidth}
\includegraphics[angle=90,width=1.\textwidth]{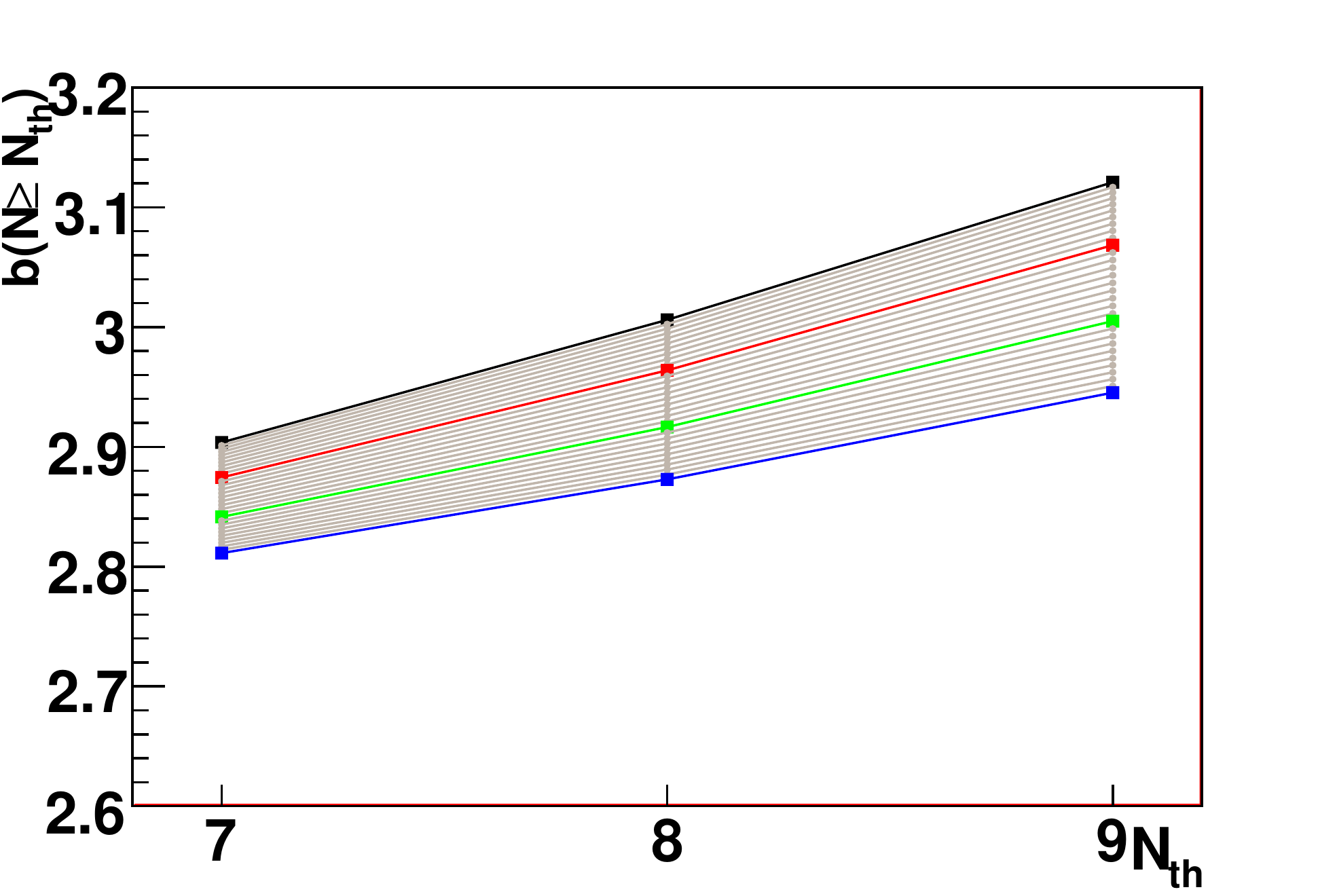}
\end{minipage}
\caption{Comparison between the model predictions with only intrinsic scatter when the minimum halo mass is taken into account (right) and when we integrate in a wider mass range (left). The circles and solid lines are the bias predictions at $z=0.5$ as a function of intrinsic scatter using the catalog created with only intrinsic scatter $\sigma_{lnM}^{true}=0.2$.}
\label{fig:bias_model_cut_nocut}
\end{figure}

\begin{figure}[htpb]
\centering
\begin{minipage}{0.49\linewidth}
\includegraphics[angle=90,width=1.\textwidth]{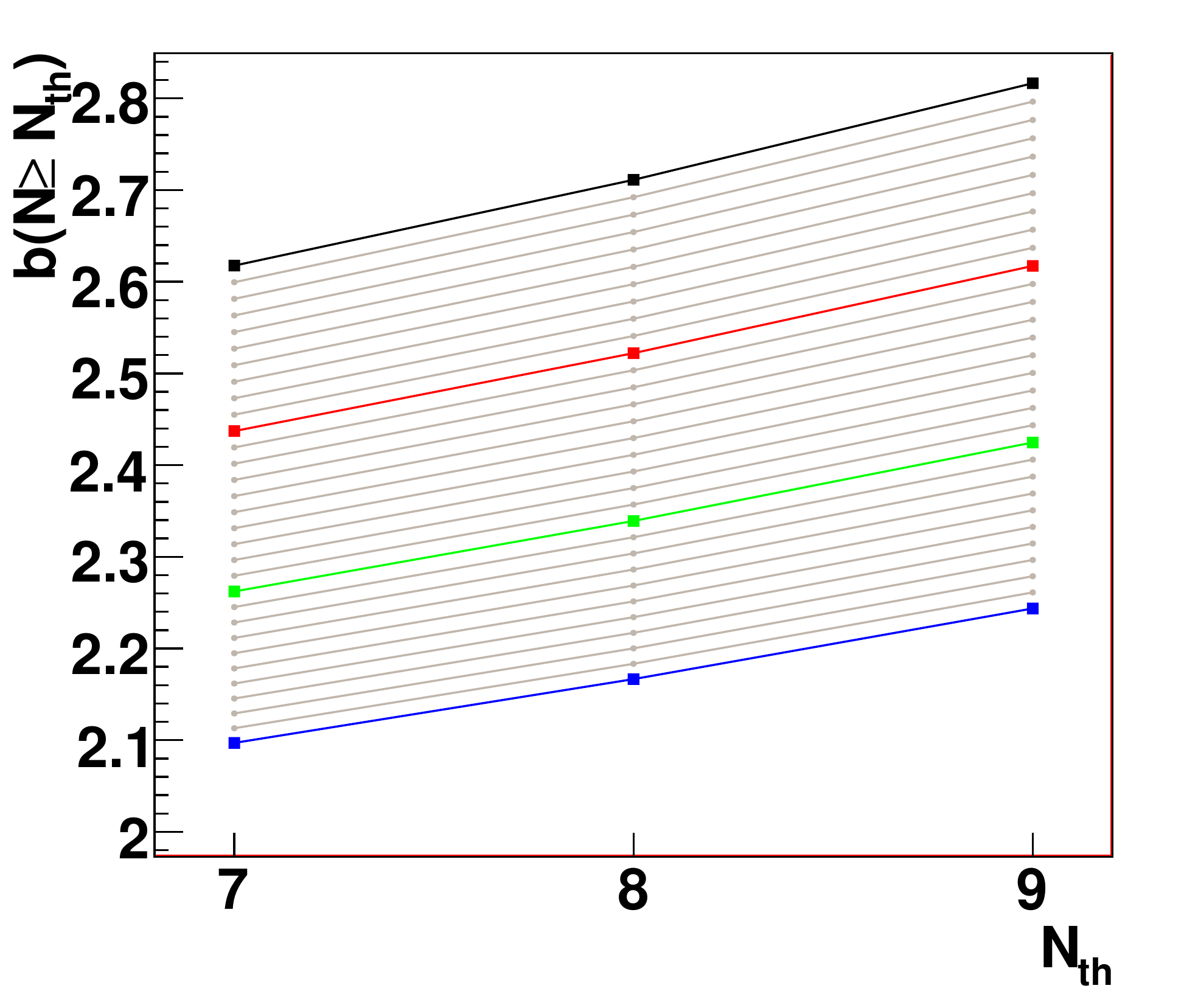}
\end{minipage}
\begin{minipage}{0.49\linewidth}
\includegraphics[angle=90,width=1.\textwidth]{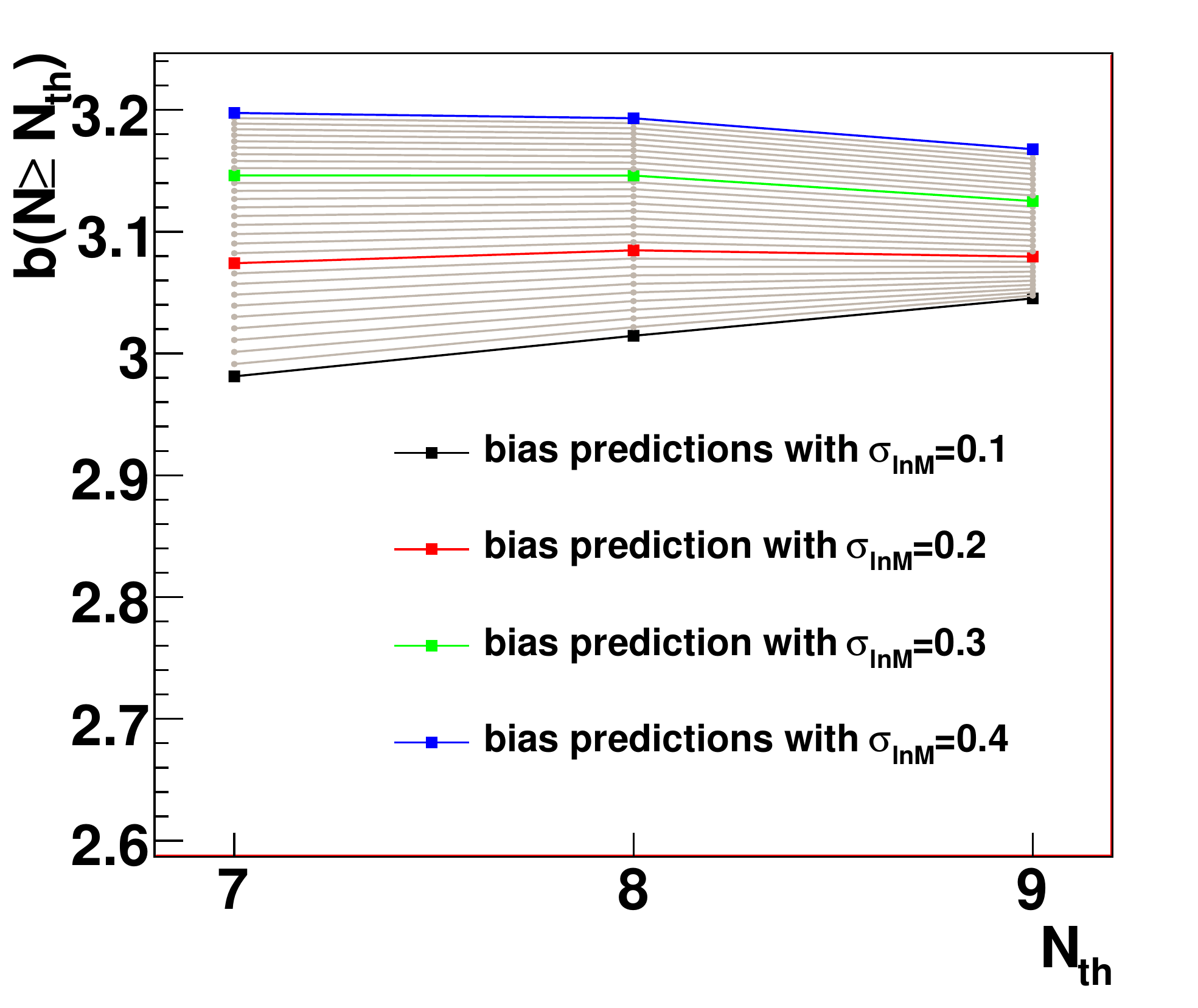}
\end{minipage}
\caption{ Comparison between the model predictions with intrinsic scatter and Poisson noise when the minimum halo mass is taken into account (right) and when we integrate in a wider mass range (left). The circles and solid lines are the bias predictions at $z=0.5$ as a function of intrinsic scatter using the catalog created with intrinsic scatter $\sigma_{lnM}^{true}=0.2$ and Poisson noise.}
\label{fig:bias_model_cut_nocut_noise}
\end{figure}

 
The black points in Figures \ref{fig:fig3} and \ref{fig:fig4} show the recovered values of the scatter $\sigma_{\ln M}$ and their $68\%$ errors $\sigma(\sigma_{\ln M})$ when the predictions only include the intrinsic scatter. We obtain the same results using \citet{1999MNRAS.308..119S} and \citet{2010ApJ...724..878T} models. Our results show that we may estimate the scatter with a standard deviation or expected error $\sigma(\sigma_{\ln M})$ $(68\% C.L)$ of  approximately $0.041$, $0.031$, $0.027$ and $0.025$ for $\sigma_{\ln M}^{true}=0.1,0.2, 0.3$ and $0.4$ respectively. The precision to measure the scatter is better at larger values because  the second derivative of the bias with the scatter is negative, $\frac{\delta^2b(N_{th},z)}{\delta ^2 \sigma_{\ln M}}$.   As we explained in Section \ref{sec:Bias_prediction}, due to the slope of the mass function that determines the excess of upscattered clusters, the larger the scatter the more the bias decreases (see Figure \ref{fig:bias_model_cut_nocut}). As an example, meanwhile the precision at $\sigma_{\ln M}^{true}=0.1$ is $40\%$, at  $\sigma_{\ln M}^{true}=0.4$ increases until $6\%$.  

Although the dominant systematic uncertainty  in our method comes from the halo mass and bias function, another source of systematics is the mass resolution of the light cone simulations. The minimum halo mass introduces a systematic that affects our richness bias model predictions especially when our observable mass is closer to the minimum. For the DES volume, we cannot choose observable thresholds further from this minimum with enough clusters to avoid this systematic. Figure \ref{fig:bias_model_cut_nocut} shows the comparison between the richness bias predictions when we integrate taking into account the cut in mass and when we integrate in a wider range. In the first case, since we are removing halos from the area where there is an excess of clusters, 
the decreasing slope of the bias with the scatter is lower  than when we don't remove them. Thus, we will loose precision to recover the scatter as the results show. Moreover, the larger  the scatter the larger the disagreement between the two cases and the minimum halo mass systematic is more significant.  As an example, when $\sigma_{\ln M}^{true}=0.4$ the precision is reduced by half, or the expected error increases from  $\sigma(\sigma_{\ln M})=0.012$ to $\sigma(\sigma_{\ln M})=0.025$ $(68\%$ $C.L.)$. 


 
After we studied the forecast constraints with a constant scatter, we add the Poisson noise in our richness bias model. We repeat the likelihood calculation to study the precision that can be reached and compare the results with the previous case. The red points in Figures \ref{fig:fig3} and \ref{fig:fig4} show the recovered values of the scatter $\sigma_{\ln M}$ and their $68\%$ errors $\sigma(\sigma_{\ln M})$ for the fiducial values $\sigma_{\ln M}^{true}$. The results show that we may estimate the scatter  with a standard deviation or expected error $\sigma(\sigma_{\ln M})$ $(68\% C.L)$  of  approximately $0.018$, $0.024$, $0.048$ and $0.05$ for $\sigma_{\ln M}^{true}=0.1,0.2, 0.3$ and $0.4$ respectively.  We conclude that when the Poisson noise is included, the precision increases considerably at the lowest scatter values $\sigma_{\ln M}^{true}=0.1$. It increases from $41 \%$ to $18\%$. However, although the precision increases smoothly for $\sigma_{\ln M}^{true}=0.2$ , it decreases by half at larger values.

The effect of adding the Poisson noise in the bias predictions is shown in Figure \ref{fig:bias_model_cut_nocut_noise}.  
The excess of upscatter versus downscatter around the observable threshold will be larger and it will be more significant in the low-mass range  at lower scatter values when we add this noise. If we would not remove the halos where there is the excess, the variation of the bias with the scatter will be constant. Therefore, the precision would be very high and  constant for all the scatter values,  $\sigma(\sigma_{\ln M})\sim 0.01$.  
However, when we remove halos where the bulk of the values lies,  the bias will increase with the scatter instead of decrease specially when our observable is closer to the minimum halo mass. Moreover, the lower the scatter, the more the bias increases when we are close to the minimum halo mass. 
We conclude that due to the minimum halo mass, we loose precision to recover the scatter  especially at larger values of the scatter. 
When $\sigma_{\ln M}^{true}=0.1$ the precision decreases from $10\%$ to $18\%$ or the expected error increases from  $\sigma(\sigma_{\ln M})=0.01$ to $\sigma(\sigma_{\ln M})\sim0.02$ $(68\%$ $C.L.)$. Meanwhile, when $\sigma_{\ln M}^{true}=0.4$ the precision decreases from $2.5\%$ to $12.5\%$ , or the expected error increases from  $\sigma(\sigma_{\ln M})=0.01$ to $\sigma(\sigma_{\ln M})=0.05$ $(68\%$ $C.L.)$.

\begin{figure}
\centering
\includegraphics[scale=0.7,angle=90]{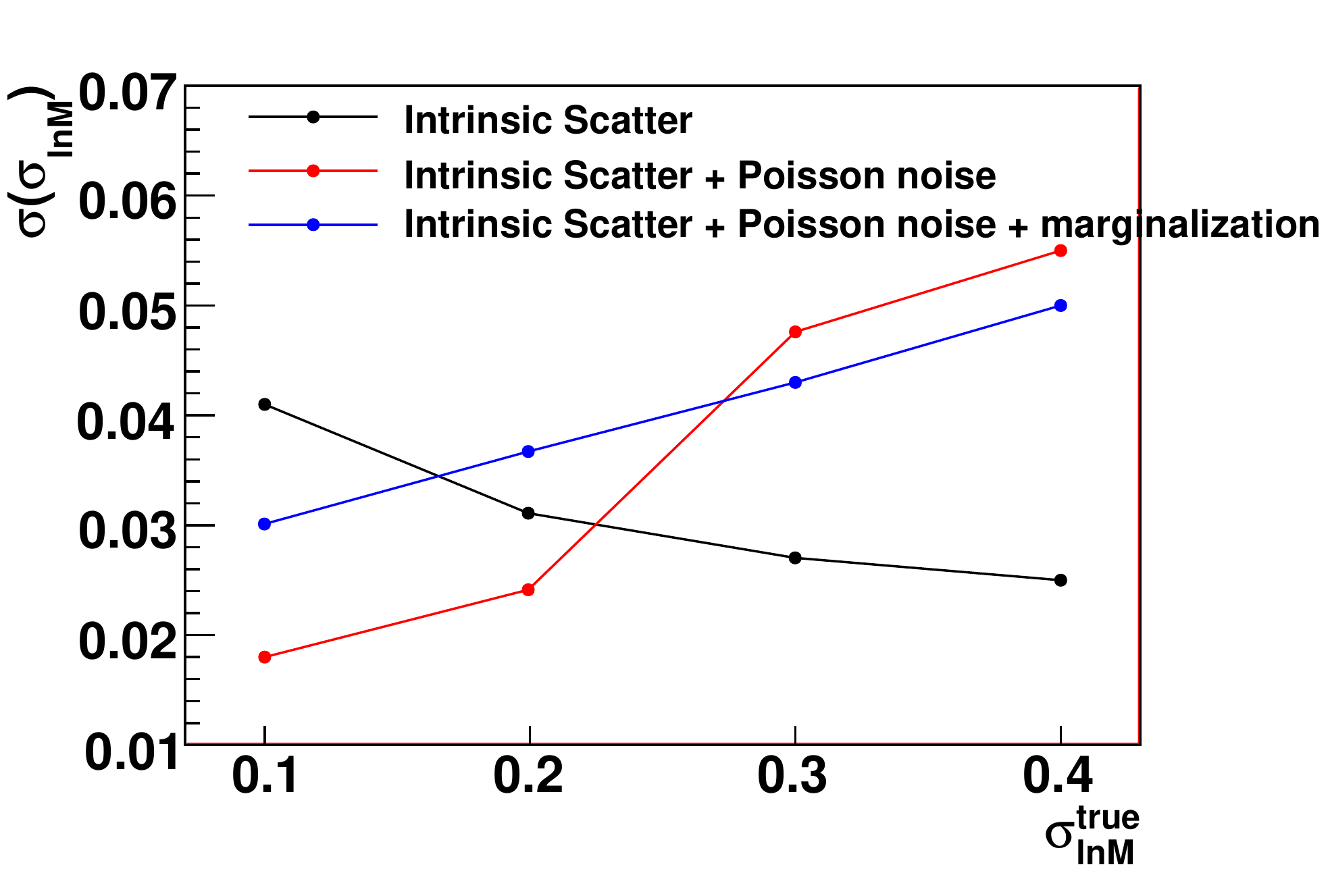}
\caption{Blue dots and lines are the expected errors ($68\% C.L.$) for three true scatter values, $\sigma_{lnM}^{true}$ when we marginalized over the mass-richness relation parameters $\alpha_{M|N}$ and $B$. The two terms in the scatter are included. The comparison to the precision when we assume we know perfectly these parameters is also shown.}
\label{fig:figmar}
\end{figure}

After we evaluate the likelihood function assuming that we know perfectly the mean mass-richness relation, we assume Gaussian distributed errors for these parameters and marginalize over them using Equation \ref{eq:marginalize}. These external priors are given in Section \ref{sec:m-N}.  Figure \ref{fig:figmar} shows the $68\%$ errors $\sigma(\sigma_{\ln M})$ for the fiducial values $\sigma_{\ln M}^{true}$ and the comparison with the constraints when we do not marginalize. Our results show that when we marginalize we loose $\sim50\%$ of  precision at  lower intrinsic scatter values $\sigma_{\ln M}^{true}=0.1$ and $0.2$; meanwhile, at the highest values  a difference of $8\%$ is not so significant. 


\subsection{Results in simulations}
After the forecast, we study how well our method can constrain the scatter of the scaling relation using the simulated cluster catalogs. Here we add the uncertainty in the theoretical models when we compare with simulations.

We perform a likelihood calculation comparing the bias prediction with the measurements for the three cluster catalogs created. 
 We divide the catalog in 6 redshift bins of width $\Delta z=0.2$ and make cuts in richness. 
First, we assume we know the mean mass-richness relation parameters $\alpha_{M|N}$ and $B_{M|N}$ and the $\Lambda CDM$ cosmological parameters of our simulations.  As in previous section, we start studying the constraints when the distribution of clusters only includes the intrinsic scatter. Then we include the Poisson noise.

\begin{figure}
\centering
\begin{minipage}{1\linewidth}
\includegraphics[scale=0.5,angle=90]{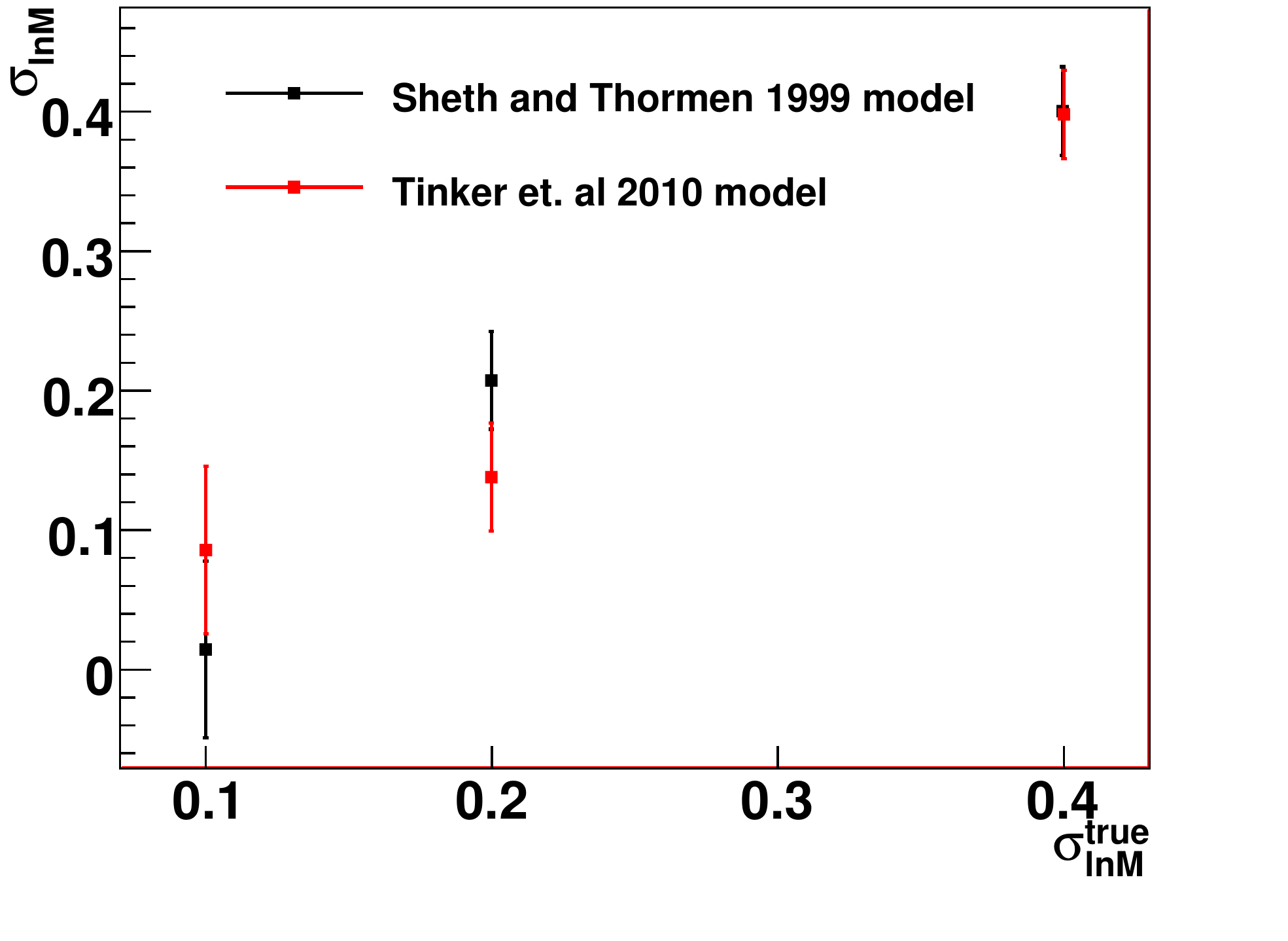}
\caption{Recovered values of the scatter, $\sigma_{lnM}$, for the three catalogs created with $\sigma_{lnM}^{true}$ when we use the Sheth and Thormen (1999) (black dots) and \citet{2010ApJ...724..878T} (red dots) models. The expected errors ($68 \% C.L$) are also shown. }
\label{fig:fig5}
\end{minipage}
\begin{minipage}{1\linewidth}
\includegraphics[scale=0.5,angle=90]{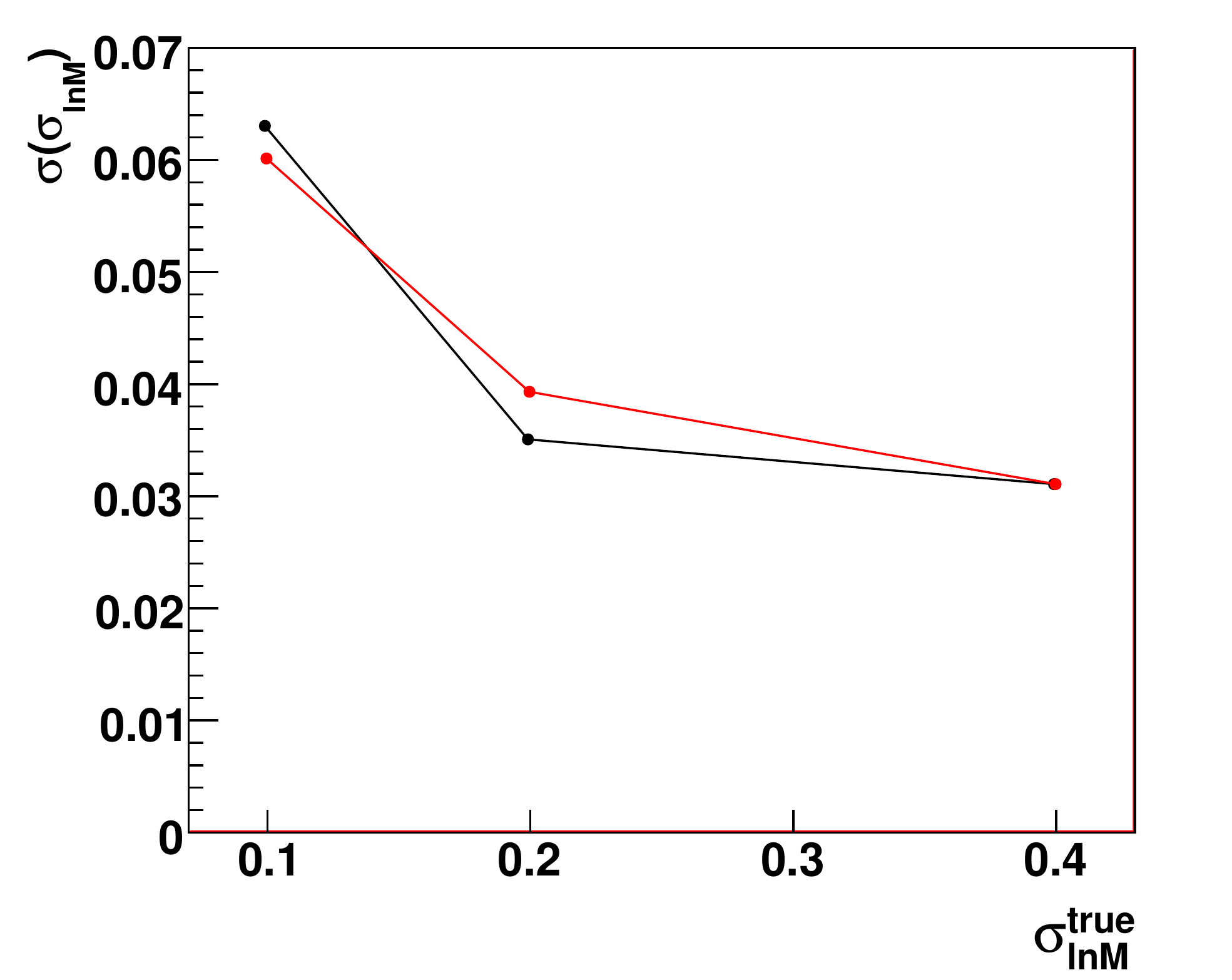}
\caption{Expected errors ($68\% C.L.$) for three true scatter values, $\sigma_{lnM}^{true}$ when we use the Sheth and Thormen (1999) (black dots) and \citet{2010ApJ...724..878T} (red dots) models.}
\label{fig:fig6}
\end{minipage}
\end{figure}

Figures  \ref{fig:fig5}  and \ref{fig:fig6} show the recovered value of the intrinsic scatter and their $68\%$(C.L) errors for the two theoretical models when the scatter is constant. As predicted by the forecast, at the largest scatter value, $\sigma_{\ln M}^{true}=0.4$,  we obtain the best precision ($8\%$) and accuracy and the two halo bias prescriptions agree, $\sigma_{\ln M}=0.399 \pm 0.031$ $(68\%$ $C.L.)$.  However, for lower values there is a discrepancy between them. When the scatter is $\sigma_{\ln M}^{true}=0.1$ we cannot recover the true value in any case while we recover it for $\sigma_{\ln M}^{true}=0.2$ using the \citet{1999MNRAS.308..119S} prescription. In this case the result is  $\sigma_{\ln M}=0.206 \pm 0.035$ $(68\%$ $C.L.)$.

Apart from the redshift uncertainty that we will study soon, we conclude that the theoretical model is the main systematic of this method. As we mention in Section \ref{sec:mf}, we postpone a careful calibration of the mass function and bias parameter of the Tinker model in our simulations to study better the difference with the Sheth \&Tormen (1999) predictions.

After we studied the constrains when the scatter is constant, we add the Poisson noise in both simulations and the theoretical models.  In this case, we only use the \citet{1999MNRAS.308..119S} prescription because we obtain better accuracy after  the $p$ and $q$ parameters calibration. 

The results show that the scatter can only be recovered with good accuracy and a precision of $18\%$ at the lowest value. We obtain that the mean and the  $68\%$(C.L) errors is $\sigma_{\ln M}=0.091 \pm 0.020$ when the intrinsic scatter is $\sigma_{\ln M}^{true}=0.1$.   
When $\sigma_{\ln M}^{true}=0.2$ the method is having difficulty. The accuracy is  $\sigma_{\ln M}-\sigma_{\ln M}^{true} \ge 0.05$, so that the dark energy parameters will be significantly biased and the noise will be observationally relevant.


Figure \ref{fig:bias_model_cut_nocut} and \ref{fig:bias_model_cut_nocut_noise} show the comparison between model predictions when the minimum halo mass is taken into account and when we integrate in a wider range. When we add the Poisson noise the difference between models are larger than when we only consider the intrinsic scatter.  When the minimum halo mass is close to the observable threshold the value of the bias and its variation with the scatter and richness changes considerably. Apart from this systematic, here we add the uncertainty in the dark matter halo bias and mass function model.  Therefore, the uncertainty in the model predictions will be larger specially for the larger scatter values and  largest richness threshold selected.

Finally, we repeat the likelihood calculation and marginalize over the forecast errors for the parameters of the mass-richness relation calibrated in the maxBCG cluster catalog. As predicted by the forecast, when we take into account these errors, our method loose precision at the lowest scatter values.  Using only the first cut in richness $N_{th}\ge7$ we obtain that $\sigma_{\ln M}=0.1 \pm 0.03$ $(68\%$ $C.L.)$ when  $\sigma_{\ln M}^{true}=0.1$. Our conclusion is that using our technique in a $5,000$ $deg^2$ volume survey we will be able to constrain the scatter at suitable precision and accuracy when $\sigma_{\ln M}\sim 0.1$. 

We want to see if our method will be precise enough for the estimated scatter value in DES. \citet{2012ApJ...746..178R} provide a rough calibration of the mass-richness relation using the redMaPPer cluster finder with maxBCG clusters. They give an estimation  of the scatter which can be used as the expected value for the DES survey. 
Early results from the DES SVA data given by \citet{2015MNRAS.449.2219M} and \citet{2015MNRAS.454.2305S}  also give an estimation of this quantity.  Using these studies, we estimate that the intrinsic scatter of the mass-richness relation for the DES survey using redMaPPer is $\sigma_{lnM}$  $\sim 0.18-0.3$ depending of the richness  although further work benefiting from a larger region will improve the constraints. 
As our results show, we conclude that the DES volume is insufficient to recover the expected scatter for the RedMaPPer cluster with this method with enough accuracy and precision.
\section{Discussion and conclusions}
In this paper, we aim to explore the capability of the spatial correlation function in photometric surveys to reduce the uncertainty in the intrinsic scatter of the mass-richness relation. We assigned richness to dark matter halos with an effective scatter in N-body simulations and measure the bias using the cluster correlation function.  First, we assume a model with constant scatter and then we add the Poisson variance. By carrying out a likelihood analysis, we forecast the precision to measure the intrinsic  scatter using a light cone simulation of $5,000$ $deg^2$ up to $z\sim 1.2$, similar to the ongoing DES project. 

Our conclusions are that the new method  works better at lower intrinsic scatter values. 
 The value $\sigma_{\ln M}=0.1$ can be recovered with a precision of $30\%$ and the  accuracy  $\sigma_{\ln M}-\sigma_{\ln M}^{true}$ is $\le 0.05$. However, we cannot recover the expected value of the RedMaPPer cluster catalog at a suitable accuracy and precision because the DES volume is insufficient. We can not choose an observable threshold further from the minimum halo mass with enough clusters. 
Note that for a cluster cosmology experiment  instead of the mass resolution, the mass systematic would be the minimum observable richness. 

Although our method will be more precise in future cluster surveys, the study of the clustering of clusters and the implementation of this new technique can be done in the real DES cluster catalogs that are available. This analysis can be used as a training for future surveys  to reduce the uncertainty in cosmological parameters coming from an uncertainty in the mass-observable parameters.

Our new technique can also be used as a cross-check method for other ways of estimating the scatter using direct methods with a small subset of clusters. In \citet{2015MNRAS.454.2305S}, they cross-match the SPT catalog with the optically selected clusters of the DES Science Verification Area (SVA), the  redMaPPer and VT cluster catalog. Although this is a very promising technique to use, one of the drawbacks is that it is limited to calibrate SPT-SZE clusters.  In addition, one of the advantages of our method is that it only uses optical clusters and we could measure the scatter in a broader mass range than the SZE clusters.
 
In a near future, the mass-richness relation will be also calibrated in the DES cluster catalog with other mass proxies such as stacked weak lensing shear. Then we can use this relation as a prior on the mass-richness relation parameters as a function of redshift instead of the one we use from maxBCG clusters. We expect that the errors of the mass-richness parameters will be reduced to improve our results when we marginalize.
  
 Apart from the minimum halo mass, the main systematic error we have found is the uncertainty in the halo mass function and bias. In addition, the difference between the \citet{1999MNRAS.308..119S} and  \citet{2010ApJ...724..878T} bias prescriptions is also systematic error. 
A next step is to perform a calibration of the Tinker parameters in  the simulations we used and see if there is still this difference. 

 In this work we ignore the effect of the uncertainty of the redshift of the clusters. Although one advantage is that this error is lower than in galaxies, it affects the three dimensional correlation function. It is effect is a smearing of the acoustic peak \citep{2009ApJ...692..265E} and a relative damping of power on small scales that reduces the bias.  We postpone a careful study of how this systematic error will affect the bias measurement and the precision of the scatter measurements. This will allow us to use the spatial correlation function (3D) and compare the results with the angular correlation function, $\omega(\theta)$. 

There is also the possibility that the scatter evolves with redshift $\sigma_{\ln M}(z)$ although in this work we have assumed that it is constant in the six redshift slices with $\Delta z=0.2$.  In previous studies the scatter increases with redshift. \citet{2010PhRvD..81h3509C} parameterized this evolution using a cubic polynomial and studied the impact of the parameters of the scatter evolution on the mass function. 
We also postpone a careful study of the effect of the scatter evolution on the bias predictions. We forecast that our precision to measure the scatter will change in all the redshift bins  if we compare with the results when the intrinsic scatter is constant, $\sigma_{\ln M}(z)=\sigma_{\ln M}(z=0)$. However, we need to quantify the change in the intrinsic scatter with redshift to study in detail the impact in our precision.

For future photometric cluster surveys with larger area such as LSST and Euclid, we forecast higher precision to constrain the scatter at the level in which is having difficulty with a DES volume. Since LSST will image $20,000$ $deg^2$, we expect the statistical errors will be reduced at high mass and redshift, because the number of clusters will increase considerably.  In a wider area, we may choose richness cuts where the observable threshold is further from the minimum mass limit.  With this, we forecast higher precision for all the scatter values. 

\begin{figure}
\centering
\includegraphics[scale=0.7,angle=90]{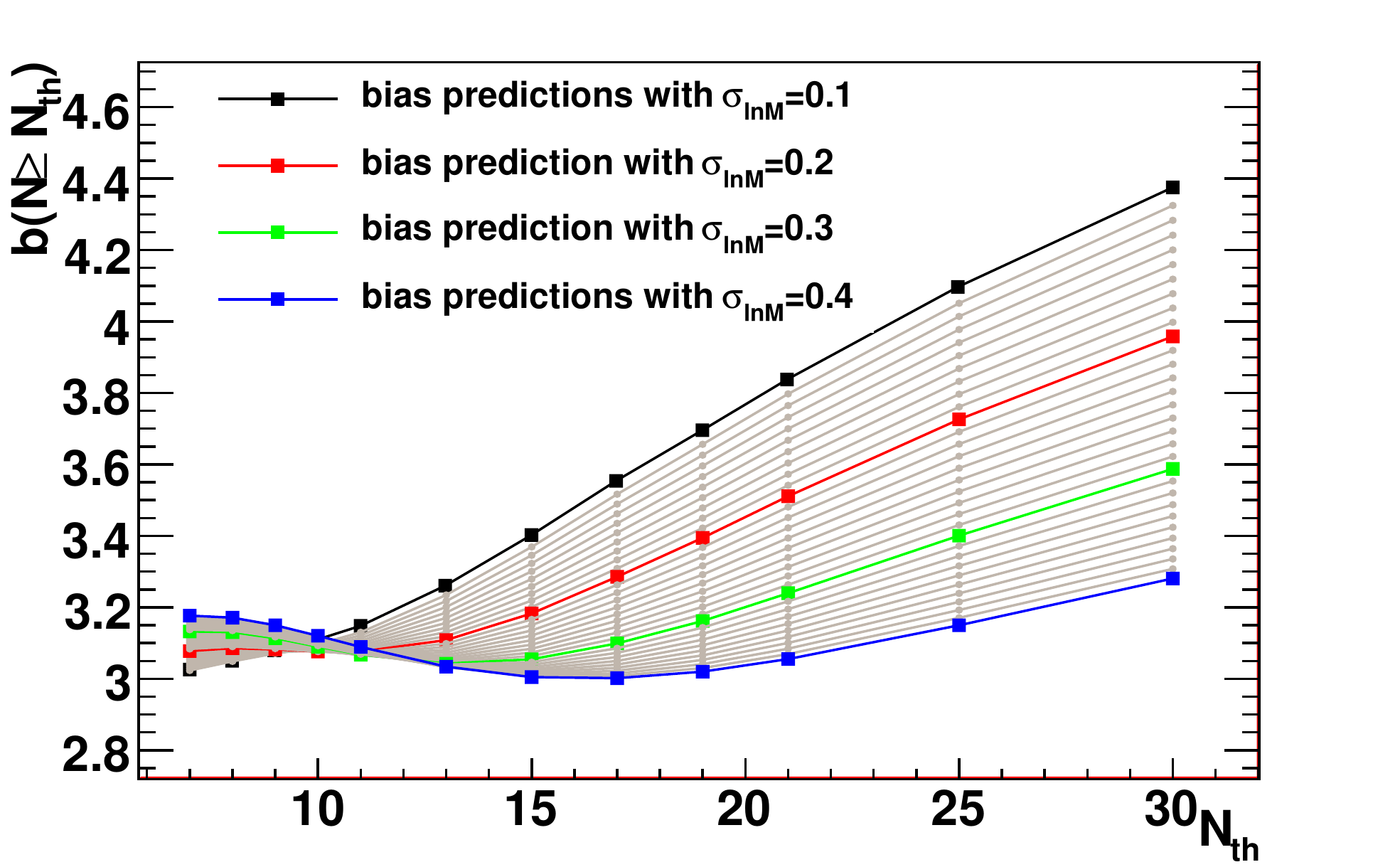}
\caption{The circles and solid lines are the bias predictions at $z=0.5$ as a function of intrinsic scatter using the catalog created with intrinsic scatter $\sigma_{lnM}^{true}=0.2$ and Poisson noise. Here we can make cuts in richness until N=30 to see how the predictions changes when the observable threshold is far from the minimum mass.}
\label{fig:fig9}
\end{figure}

Figure \ref{fig:fig9} shows the bias predictions as a function of intrinsic scatter when we make cuts in richness at larger values than DES. As an example, if we make richness cuts at $N=13,15, 17$ and $19$ we forecast that  the precision to measure $\sigma_{\ln M}^{true}=0.2$ will increase twice. The intrinsic scatter will be constrained to be $\sigma_{\ln M}=0.2\pm0.012$ and our precision will increase from $12\%$ to $6\%$. 
With this, we forecast higher precision to measure the scatter of the mass-richness relation including the values where our method is having difficulty with a DES volume.

In summary, we conclude that using only large optical photometric cluster surveys the new method proposed is a promising method that can be use as a cross-check method for other methods that use multi-wavelength observations. 

\acknowledgments
We greatly appreciate the support received from the collaborative work Julia Campa undertook with Martin Makler,  Mariana Penna and Marc Manera, as we worked together on the theoretical predictions of the halo mass function and bias. We are grateful to Jim Annis, Tom Diehl, Marcelle Soares, Brian Nord, Liz Buckley-Geer, David Finley and Josh Frieman for useful discussions and communications. Thanks again to David Finley for laboriously correcting the grammar and languages mistakes, and making suggestions as to what should be explained more.  Julia Campa thanks all participants of the Experimental Astrophysics Group meetings at the Fermilab Center for Particle Astrophysics. Thanks to Eduardo Rozo, Eli Rykoff, Joe Mohr, Cristopher Miller, Risa Wechsler and all the participants of the DES cluster working group. Thanks to Michael Busha and Risa Wechsler for providing us the DESv1.02 halo mock catalog light cone survey based on HVS simulations.  Julia Campa gratefully acknowledges the funding sources that made this work possible. Much of this work was supported by the Astroparticle Physics Division at Centro de Investigaciones Medioambientales y Energ\'eticas  (CIEMAT). This work was partially completed at Fermilab. Julia Campa acknowledges the financial suppport provided by the Particle Physics Division at Fermilab and the 
 invitations to work at the Fermilab Center for Particle Astrophysics (FCPA). 

\appendix

\clearpage



\clearpage

\clearpage

\clearpage

\clearpage

\end{document}